\documentclass[a4paper]{jpconf}
\usepackage{graphicx}
\usepackage{color}
\usepackage{ulem}

\newcommand{\csout}[1]{{\color{blue}\sout{#1}}}
\newcommand{\crep}[2]{{\color{blue}\sout{#1}}{\color{red}#2}}
\newcommand{\cin}[1]{{\color{red}#1}}
\newcommand{\csoutt}[1]{{\color{blue}\sout{#1}}}
\newcommand{\cinn}[1]{{\color{red}#1}}
\newcommand{\crepp}[2]{{\color{blue}\sout{#1}}{\color{red}#2}}

\renewcommand{\csout}[1]{{}}
\renewcommand{\crep}[2]{#2}
\renewcommand{\cin}[1]{{#1}}
\renewcommand{\crepp}[2]{#2}
\renewcommand{\cinn}[1]{{#1}}
\renewcommand{\csoutt}[1]{{}}

\begin{document}

\title{Bulk-edge correspondence in topological transport and pumping}

\author{Ken-Ichiro Imura$^{1}$, Yukinori Yoshimura$^{1,2}$, Takahiro Fukui$^3$, Yasuhiro Hatsugai$^{4}$}
%\author{Ken-Ichiro Imura$^{1,2}$, Yukinori Yoshimura$^{1,3,4}$, Takahiro Fukui$^5$, Yasuhiro Hatsugai$^{6,7}$}

\address{$^1$Department of Quantum Matter, AdSM, Hiroshima University, 739-8530, Japan}
%\address{$^2$Center for Emergent Condensed-Matter Physics, Department of Physics, Hiroshima University, 739-8526, Japan}
\address{$^2$MathAM-OIL, AIST \& Tohoku University, Sendai 980-8577, Japan}
%\address{$^4$AIST}
\address{$^3$Department of Physics, Ibaraki University, Mito 310-8512, Japan}
\address{$^4$Institute of Physics, University of Tsukuba, Tsukuba 305-8571, Japan}
%\address{$^7$BEC}

\ead{imura@hiroshima-u.ac.jp}

\begin{abstract}
The bulk-edge correspondence (BEC) refers to a one-to-one relation
between the bulk and edge properties ubiquitous in topologically nontrivial systems.
Depending on the setup, 
BEC manifests in different forms and govern the spectral and transport properties of topological insulators and semimetals. 
%Here, we focus on the case of　topological pumping and 
%clarify compensating roles of the bulk and the edges of the system.
\cin{
Although the
topological pump is theoretically
old, BEC in the pump has been established just recently \cite{HF} motivated by the 
state-of-the-art experiments using cold atoms
\cite{kyoto,lohse}.
The center of mass (CM) of a system with boundaries
shows a sequence of quantized jumps in the adiabatic limit
associated with the edge states.
Although the bulk is adiabatic, the edge is inevitably non-adiabatic in the experimental setup
or in any numerical simulations.
Still the pumped charge is quantized and carried by the bulk. 
Its quantization is guaranteed by a compensation between the bulk and edges.
%the center of mass
We show that in the presence of disorder
the pumped charge continues to be quantized despite the appearance of
non-quantized jumps.
}
%We have previously focused on the stability of surface states against lattice imperfections [2], cases of weak topological phases [3], and of Weyl semimetal thin films [4]. Quantization of pumped charge and spin is another manifestation of the nontrivial topological properties in the bulk. To quantify topological pumping, time evolution of the initial ground state is to be considered in the adiabatic limit [5,6]. Here, using the prescription of Ref. [6], we study the robustness of topological pumping against (on-site) disorder. FIG. 1 shows time dependence of the system's polarization, indicating that the pumped charge is unchanged in spite of the irregularities due to disorder. This “snapshot picture” (an approach from the adiabatic limit) reveals the role of edge states in topological pumping, providing also an interesting twist on the BEC picture.
%In topological pumping, the adiabaticity of the system plays also an important role. 
\end{abstract}

\section{Introduction}
%\subsection{Historical background}

Recently, the role of topology is often highlighted in condensed-matter physics.
This new trend in condensed-matter dates back to a few papers published %more than 
three decades ago.
One of them is the so-called TKNN paper
\cite{TKNN},
in which topological interpretation was given to
quantum Hall effect (QHE),
which had been discovered experimentally a few years earlier.
The concept of topological \cin{pump} is a variant of the idea of TKNN,
applied to a temporal evolution of the system, instead of
to the Brillouin zone.
In Ref. \cite{TKNN},
quantization of the Hall plateaus was given an interpretation
as manifestation of an underlying topological order,
which encodes a nontrivial phase property of the bulk wave function.
Apparently,
Thouless, one of the authors of Ref. \cite{TKNN} had the idea of applying the same scenario
almost at the same time to the system of \cin{adiabatic pump}
\cite{TP}.
Yet \csout{, to our knowledge}
no clear experimental demonstration of this idea had been reported
until in 2015
two experimental groups embodied
the idea of
topological \cin{pump} {\it a la} Ref. \cite{TP} in the system of cold atoms;
\cite{kyoto,lohse}
not to mention that
the original proposal was based on an electronic system.
One of the issues
yet to be investigated in the experimental studies
is on the role of disorder in pumping,
which we focus on in this paper.

\section{Contribution of the bulk vs. the edge: the bulk-edge correspondence}

A natural way to quantify pumping is to keep truck of a temporal evolution 
of the \csout{bulk} \cin{many body} wave function.
In case of topological \cin{pump}
this temporal evolution is adiabatic so that 
following the snapshot $\bar{x} (t)$ of the center of mass (CM) 
of the ground state
\cin{
\begin{equation} 
\bar{x} (t) = \sum_i x_i n_i
\end{equation} 
is enough to describe the pumping where $n_i$ is a many body particle number at  $x_i$
\cite{HF}.
Here taking a rescaling as
\begin{equation}
  x_i = \frac {i-i_0}{L} ,\quad i_0=\frac {L}{2} , \ (i=1,\cdots, L)
  \label{eq:scaling}
\end{equation}
is essential ($L$ is the system size).}
To demonstrate this
we consider a $1+1$-dimensional (i.e., 1 spatial $+$ 1 temporal dimensions) model;
a one-dimensional (1D) model with an (effective) time-dependent potential
[see Eq. (\ref{ham})].
For the practical numerical simulation
we employ the pump version of the so-called Harper, or Aubry-Andre model.
\cite{HF}
Then, we impose the boundary condition such that
the system is periodic in time,
while it is open 
i.e., with boundaries
\footnote{may sound paradoxical, but this seems to be the standard terminology in the field.}
in the space direction.
This means that the snapshots of the CM is \cinn{well-defined and } periodic in time:
\begin{equation}
\label{periodic}
\bar{x} (t_0 + T) - \bar{x} (t_0) = 0,
\end{equation}
where $t_0$: initial time, $T$: pumping cycle.
This might seem to imply that
it is impossible to quantify pumping in this way.
Let us recall, however,
in the typical situation we consider in topological \cin{pump},
the (one-body) spectrum of the system
is characterized by
the existence of edge modes traversing \csout{(in reciprocal space)}
the bulk energy gap
[see FIG. 1 (a)].
Therefore, the Fermi energy
set typically in the gap intersects with such edge modes
in the course of the time evolution.
Then,
if we consider an evolution of CM of the (many-body) ground state
$\bar{x}(t)$,
it has two distinct parts which can be readily separable;
patches of continuous curves and discrete jumps
[see FIG. 1 (b)].
The jumps are necesarilly associated with edge modes {\it in the clean limit}, 
and their magnitudes $\Delta\bar{x}_{\rm jump}$ are always
quantized to be {\it half integral}:
$\Delta\bar{x}_{\rm jump} = \pm 1/2$.
\cite{HF}

In topological insulators and related systems
a one-to-one relation can be established between the appearance of edge/surface modes and
the topological non-triviality in the bulk.
The bulk-edge correspondence (BEC) refers to this one-to-one relation.
\cite{BEC}
Here, in our choice of the boundary condition
(open in one and periodic in the other),
which was also the case in the so-called Laughlin's argument,
\cite{Laughlin}
the effects of the bulk and edges are %somewhat {\it mixed}.
{\it superposed} and {\it interconnected}.
The evolution of the CM in continuous patches
is due to the bulk,
while the jumps are due to the edge modes.
To quantify pumping
and reveal
the compensating roles of the bulk and the edge,
we attempt to separate the effect of the bulk and that of the edges.
%Since the appearance of edge modes is subjective to the choice of boundary conditions, let us first focus on what the bulk does to the CM.

To concretize this BEC %procedure
we reconnect the discrete patches of the CM curve
by eliminating the discontinuities,
and form a continuous CM curve over the cycle.
Note that the resulting
continuous curve is no longer periodic in time but it acquires
a net gain (or loss) $\Delta\bar{x}_{\rm net}$ per cycle.
One can interpret this  $\Delta\bar{x}_{\rm net}$ as the net pumped charge,
transported {\it through the bulk}.
\cinn{This is a polarization of the bulk.}
Since the net effect of
bulk and edge contributions
cancel after a cycle [see Eq. (\ref{periodic})],
\begin{equation}
\label{net}
\Delta\bar{x}_{\rm net} = - \sum_{\{j_n\}}\Delta\bar{x}_{\rm jump}  (t_{j_n}),
\label{eq:becp}
\end{equation}
where the summation is over the jumps, i.e., discontinuities of $\bar{x} (t)$
due to the ``appearance'' or ``disappearance'' of an edge mode
in the ground state subspace.
Here, we are in a ``grand-canonical point of view,''
\cite{HF}
in which all the states below $\epsilon_F$ is occupied (in the ground state).
$\{j_n\}=\{j_1, j_2,\cdots\}$ represents a set of time slices where the jumps occur (see Sec. 3 for its precise definition).
The number $N_e$ of the occupied states below $\epsilon_F$, i.e., the number of electrons
changes by 1 at such jumps. 
Since after a complete cycle of time $T$
this number must get back to the original value,
the number of times an edge state ``appears'' must be equal to the number of
times an edge state ``disappears'' so that
the total number $N_{\rm jump}$ of jumps is {\it even}.
This immediately results in
that the pumped charge $\Delta\bar{x}_{\rm net}$ per cycle
is quantized to be an {\it integer}\cin{\cite{HF}}.

This integral quantization of $\Delta\bar{x}_{\rm net}$ has a profound mathematical meaning.
In parallel with the case of QHE
one can directly relate
$\Delta\bar{x}_{\rm net}$ to a topological (Chern) number
\cite{TP}.
In QHE, quantization of the Hall conductance $\sigma_{xy}$ was attributed to
the existence of an underlying topological number.
\cite{TKNN,Kohmoto}
Here, quantization of $\Delta\bar{x}_{\rm net}$
has the same mathematical origin.
In case of FIG. 1
more than two: $n_F\ge 2$ bands are fully occupied below the Fermi energy $\epsilon_F$,
which is set to be between the $n_F$-th and $n_F+1$-th bands.
Then, $\Delta\bar{x}_{\rm net}$ becomes the sum of all the Chern numbers associated with a filled band: 
\cinn{
  \begin{equation}
\Delta\bar{x}_{\rm net}=C(n_F),\quad C(n_F)=\sum_{n=1}^{n_F}C_n,
  \end{equation}
  }
where $C_n$ is a Chern number associated with $n$th band.
Or if one rather defines
\crepp
{$I(n_F)=\sum_{n=1}^{n_F}C_n$}
{$I(n_F)=    - \sum_{\{j_n\}}\Delta\bar{x}_{\rm jump}  (t_{j_n})$},
      then
$I(n_F)$ represents the number of (the pair of) edge modes \cinn{with suitable sign} that appear at $\epsilon_F$.
The edge quantity $I(n_F)$ is connected to the bulk quantity $C(n_F)$
through \cinn{Eq.(\ref{eq:becp}). This is the BEC relation\cite{HF,BEC,Riemann} in the
  topological pump:
  \begin{equation}
    I(n_F)=C(n_F),\quad 
    I(n_F)-I(n_F -1) =C_{n_F}.
      \end{equation}
    }
%which gives the contribution of the $n_F$th band to $\Delta\bar{x}_{\rm net}$.

\begin{figure}[h]
(a)
\includegraphics[width=20pc]{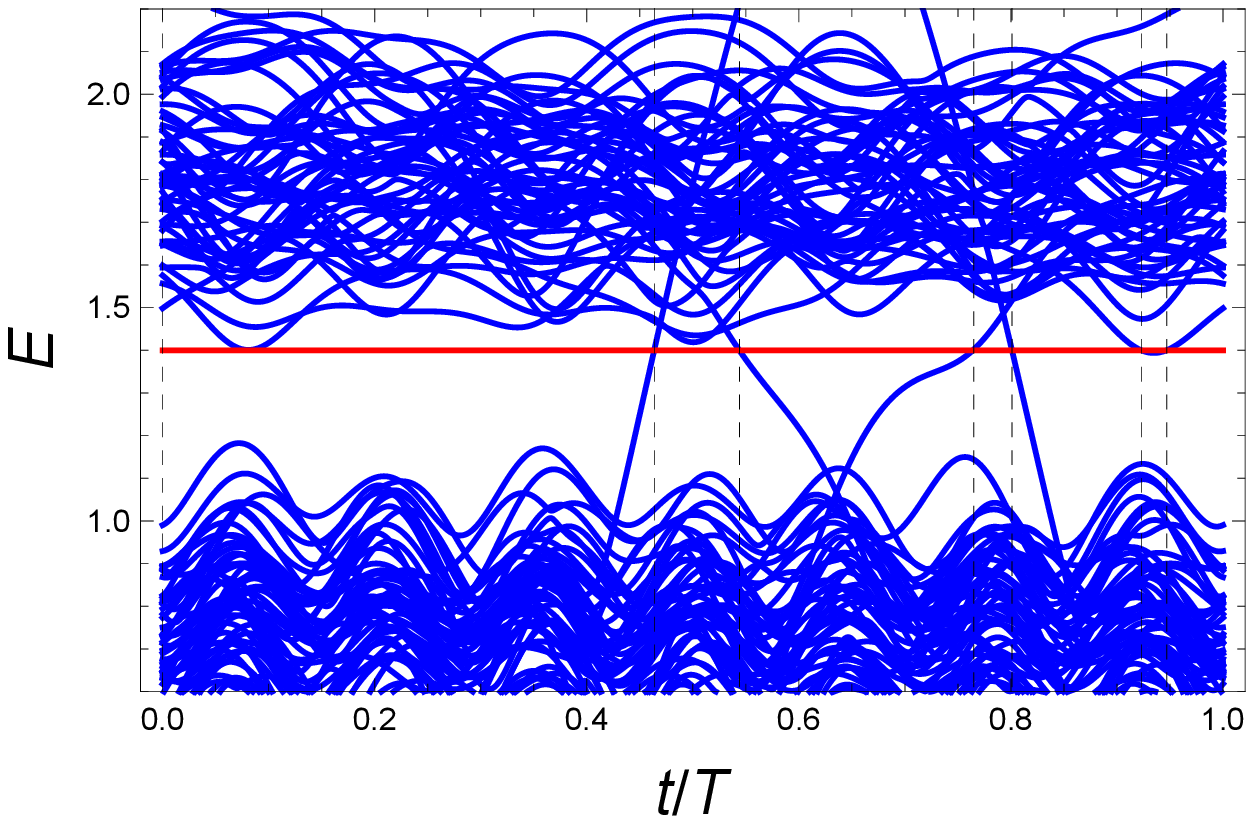}
\\
(b)
\includegraphics[width=20pc]{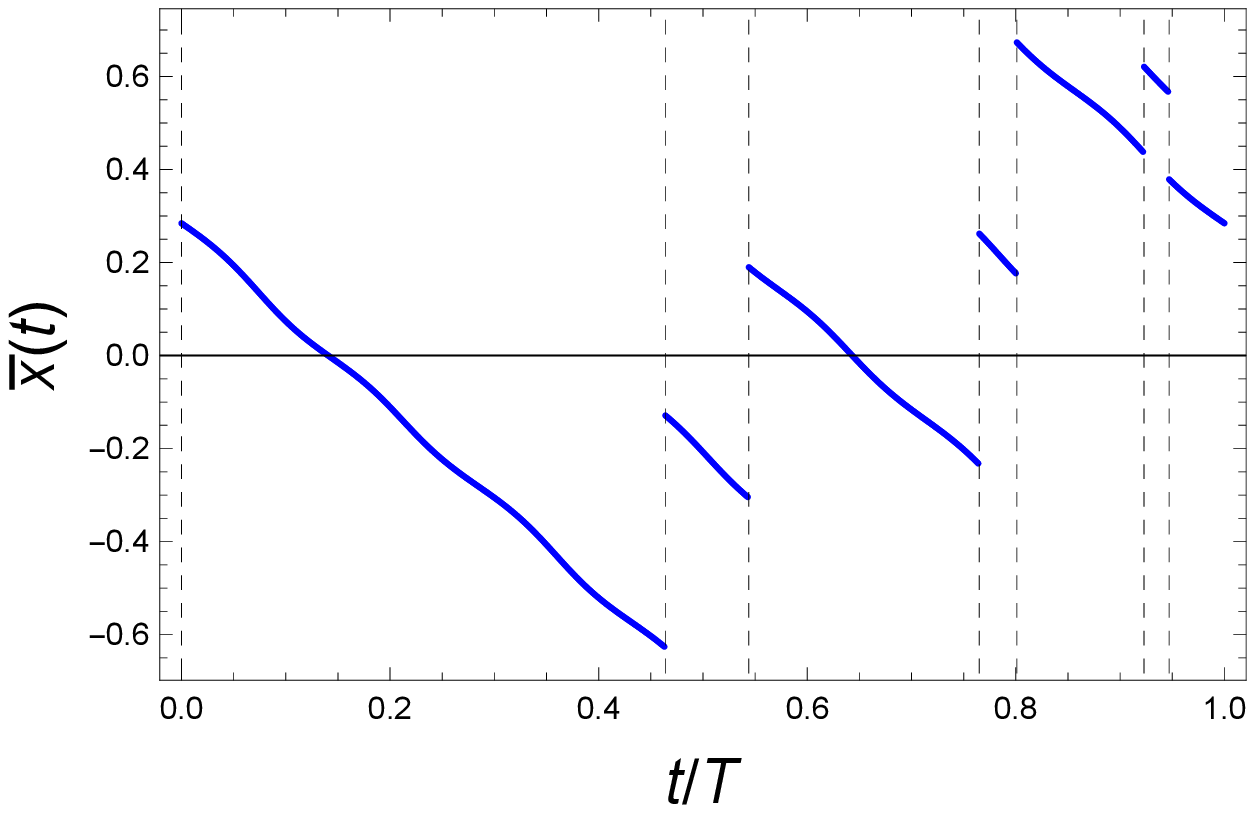}
\hspace{2pc}
\begin{minipage}[b]{14pc}\caption{\label{label}
Time evolution of (a) the snapshot spectrum $\epsilon_\alpha (t)$, and (b) the center of mass
$\bar{x}(t)$
of the ground state in case of the disordered Harper model.
In panel (b)
the magnitude of the 6 jumps are in the order of their appearance
0.497119, 0.494151, 0.494077, 0.496461, 0.182738, $-0.188663$.
Model is given in Eq. (\ref{ham}).
Parameters are specified toward the end of Sec. 3.}
\end{minipage}
\end{figure}

\section{Quantify pumping by the snapshots}

To implement the features of topological \cin{pump},
it is convenient to work on a simple theoretical model.
Even experimentally,
such an approach is proven to be useful,
e.g., in the system of cold atoms,
\cite{kyoto,lohse}
already referred to.
In Refs. 
\cite{kyoto,lohse}
the so-called Rice-Mele model was presumed,
and an effective situation in which a description by the Rice-Mele model 
has been realized experimentally in a optical lattice.
On the other hand,
Ref. \cite{HF} considers the case of Harper model,
in which situations represented by a high ($\ge 2$) Chern number
can be readily realized.
In the practical implementation we also consider this model, but
for the time being,
we can still work on a general 1D model with a time-dependent potential:
\begin{equation}
\label{ham}
H(k_y)=
%\sum_{x=-L/2}^{L/2} \bigg(
\cin{
  \sum_{i=1}^{L}
}
\bigg(
  |x_{i+1}\rangle
 t_x
\langle x_i| + 
  |x_{i}\rangle
 t_x^*
\langle x_{i+1}|  
+ |x_i\rangle \cin{
  \big(V(x_i,k_y)+W(x_i)\big) }
\langle x_i|
\bigg),
\end{equation}
where
the replacement: 
$k_y \rightarrow 2\pi t/T$
is presumed.
\cinn{\cite{HF,TP}}
In case of the Harper model the potential term becomes
\crep{
  $V(x,k_y) = 2 t_y \cos (k_y - 2\pi \phi x)$,
}
     {
       $V(x_i,k_y) = 2 t_y \cos (k_y - 2\pi \phi i)$,
     }
     where
in this original 2D representation,
$\phi$ represents the strength of a magnetic flux piercing a plaquette,
while
$\phi$ enters the 2D hopping Hamiltonian 
through \cin{Peierls} substitution.
$W(x)$ is a site random potential distributed uniformly in the range $W(x)\in [-W_0/2, W_0/2]$
\footnote{
  \cin{  This one dimensional randomness was discussed as a fictitious one in the original 2D problem\cite{Riemann}.
  It is now realized as it is in the 1D topological pump.}
  }.
Note that $W(x)$ is randomly distributed in space, while once this distribution is chosen,
it stays {\it static}.
$W_0$ specify the strength of disorder.
Eq. (\ref{ham}) is a Fourier transform of that \cin{2D-}hopping Hamiltonian.
In Eq. (\ref{ham}), $|x\rangle$ represents a Bloch state
$|x,k_y\rangle = \sum_y e^{i k_y y} |x,y\rangle$,
while we make the replacement: 
$k_y \rightarrow 2\pi t/T$,
in order to make Eq. (\ref{ham}) a 1D pump Hamiltonian.

Let us consider snapshots of such a Hamiltonian
as given in Eq. (\ref{ham}) with $k_y = 2\pi t/T$
at $t= j \Delta t$ ($j=1,2,3,\cdots$).
%where $\Delta t = T/N_t$ is the time resolution of our snapshots.
At each time slice $t_j = j \Delta t$
we diagonalize the Hamiltonian Eq. (\ref{ham})
and find eigenstates, which include both the bulk and the edge states. 
Then, we consider the ground state of the system
in which all the states below $\epsilon_F$ is occupied;
both at the edge and in the bulk.
We typically consider the case in which the Fermi energy $\epsilon_F$ is 
in the gap, since \cin{the pumped charge}  is topologically quantized in this case.
\cin{
  Since the present case is non interacting, the center of mass $\bar{x}(t_j)$
of the many body ground state is given as
\begin{equation}
\label{CM}
      \bar{x}(t_j) =
%          {\sum_{\alpha}}' x|\psi_\alpha (x,t_j)|^2,
\sum_{i=1}^L x_i n_i(t_\cinn{j}), \ n_i(t_j) =  {\sum_{\alpha}}' |\psi_\alpha (x_i,t_j)|^2,
\end{equation}
}
at different time slices $t_j$, where the summation ${\sum_{\alpha}}'$ is taken over 
all the occupied states $\alpha$ in the ground state.
$\psi_\alpha (x_i, t_j)$  represents the eigenwavefunction
corresponding to the eigenenergy $\epsilon_\alpha (t_j)$.

FIG. 1 shows the evolution of the spectrum $\epsilon_\alpha (t)$
[panel (a)]; here, the erratic behavior of the spectrum is due to the disorder potential,
and that of the center of mass $\bar{x}(t)$ in the ground state [panel (b)].
In practice,
we plot simply the spectrum $\epsilon_\alpha (t_j)$ and
$\bar{x}(t_j)$ at different time slices
to visualize their evolution.
In panel (a) one can observe that
four branches of edges modes appear and traverse the energy gap.
In panel (b)
the CM curve
$\bar{x}(t)$
shows predominantly a continuous evolution except at a few 
[actually, six in the specific case of panel (b)] discontinuities (jumps),
which occur typically
when the equi-energy line at $\epsilon_F$ intersects with either of these edge branches.
In panel (b)
this is the case at the first four jumps, while the remaining jumps are due to impurities. 
Suppose that at $t=t_{j_1}$ a ``right'' edge mode localized in the vicinity of the right edge at
\crep{$x=+L/2$}{$x=+1/2$}.
becomes {\it available} in the subspace of the ground state:
$\epsilon \leq \epsilon_F$.
Then, the number $N_e (t)=\cin{\sum_i n_i(t)}$ of occupied states increases by one at this time slice:
$N_e (t_{j_1}) - N_e (t_{j_1-1}) =1$.
Correspondingly, $\bar{x}(t)$ shows a quantized jump of $+1/2$;
i.e., $\Delta \bar{x}_{j_1} \cin{=} +1/2$.
In panel (b) this seems to happen at the second and at the fourth jump.
Generally, such a change of the occupied states $N_e (t)$ occurs at
a set of a finite number of time slices: $t=t_{j_1}, t_{j_2},\cdots$,
and there, $\bar{x}(t)$ possibly shows discontinuities.
$\{j_n\}$ in Eq. (\ref{net})
specify the set of time slices $\{j_1, j_2,\cdots\}$ 
at which the jumps occur.
In the clean limit and if $\epsilon_F$ is in the gap,
all of such intersections are associated with an edge mode, and
at each $t_{j_n}$ ($n=1,2,\cdots$)
$\bar{x} (t)$ shows a quantized jump of magnitude 
\csoutt{(ideally to be)}
%\footnote{provided that the edge mode is localized at a single site}
1/2,
\cinn{since the localization length tends to zero after rescaling  Eq.(\ref{eq:scaling})
        at the extremity of the system.}
The sign of the jump
depends on the location and the slope of the edge mode:
\cite{HF}
\begin{equation}
\Delta \bar{x}\cin{=} {1\over 2}{\rm sgn} (x_{\rm edge}) [-{\rm sgn}({\rm slope})],
\end{equation}
where
\crep
    {$x_{\rm edge}=\pm {L/2}$}
    {$x_{\rm edge}=\pm {1/2}$}
represents the location of the edge state, while
when its slope is positive (negative)
the state becomes empty (occupied) at the time slice in question.
In other words, the factor
$[-{\rm sgn}({\rm slope})]$ is a measure of the appearance/disappearance
of the state in question
in/from the ground state.

In the case of panel (b) in FIG. 1
the contribution of the first four jumps (associated with an edge state)
to the summation on the right hand side of Eq. (\ref{net}) is close to $+2$,
i.e.,
$\sum'_{\{j_n\}}\Delta\bar{x}_{\rm jump}(t_{j_n}) \cin{=} +2$,
where we used the notation $\sum'_{\{j_n\}}$ to make explicit that
only the contribution from the edge states is considered \cin{(by
  just counting the discontinuities)}.
Recall that these half-integral quantized jumps always appear {\it in pairs}.
Accordingly, the pumped charge in the bulk becomes
$\Delta\bar{x}_{\rm net} \cin{=} -2$,
which is indeed identical to the bulk topological (Chern) number.
In the situation of panel (b)
the flux $\phi$ and $\epsilon_F$ are chosen such that
\crep{$\phi = 3/7$}{$\phi=1/7$, $L=351$ and } $\epsilon_F=1.4$
with $t_x=t_y=1$.
As a result,
5 of 7 bands are fully occupied; $\epsilon_F$ is in the gap between the 5th and 6th bands.
The Chern number $C_n$ characterizing the occupied bands are \cin{$+1, +1, +1, -6, +1, +1, +1$} from the bottom to the valence band
so that they sum up to $-2$.

\section{Role of disorder: non-quantized jumps vs. quantized pumped charge}
In FIG. 1(b) one can observe that
in addition to the quantized jumps we have focused on so far,
there are additional jumps which are not quantized and appear ``trivially'' in pairs.
The fifth and sixth jumps in FIG. 1(b) fall on this category.
These additional jumps are due to impurity states.
The direction and magnitude of such jumps are such that
  \begin{equation}
\Delta \bar{x}\cin{= x_{\rm imp}}  [-{\rm sgn}({\rm slope})],
  \end{equation}
where
$ x _{\rm imp}$
\footnote{
  \cin{Due to the scaling Eq.(\ref{eq:scaling}), the position of the localized state
  is unambiguously specified as far as the localization length is finite. }
  }
represents the location (CM) of the impurity state, while
the factor $[-{\rm sgn}({\rm slope})]$
indicates
whether that appear or disappear in/from the ground state at the particular time slice.
Another implication of this factor is that
a given impurity state gives a pair of contributions 
to the summation on the right hand side of Eq. (\ref{net}) 
with the same magnitude but with opposite signs,
so that their contributions simply cancel each other.
This is contrasting to the case of half-integral quantized jumps due to edge modes
which also appear ``in pairs'' but in a different sense.
In the case of quantized jumps
the factor ${\rm sgn} (x_{\rm edge})$ allows them to evade this cancellation
and can still give a non-vanishing (though integral quantized) 
contribution to $\Delta\bar{x}_{\rm net}$.
Thanks to this cancellation,
the calculated value of  $\Delta\bar{x}_{\rm net}$ in the case of FIG. 1(b) is
$\Delta\bar{x}_{\rm net} = -1.97588$, which is close to the ideal value $-2$ in the clean limit
[in the case of two panels in FIG. 1 the strength of disorder $W_0$ is set as $W_0=1$].
This example shows that despite the appearance of non-quantized jumps
the pumped charge can still be quantized in the presence of disorder.

\section{Concluding remarks: comments on the experimental situations}
Let us summarize what we have argued so far. For the actual time evolution 
to be strictly identical to the collection of snapshots, the system must be 
adiabatic. This is the case when the pumping cycle $T$ is long enough, 
satisfying the inequality $T\gg \hbar/\Delta\epsilon$, 
where $\Delta\epsilon$ is the characteristic energy scale.
In the ``bulk regime'' in which $\bar{x}(t)$ shows a continuous evolution,
$\Delta\epsilon = \epsilon_g$ (scale of the bulk energy gap), i.e.,
the adiabaticity is controlled by $\epsilon_g$.
In the vicinity of the jumps, on contrary, or in the ``edge regime'' 
$\Delta\epsilon_{\rm edge}\rightarrow 0$, since the edge is gapless,
so that the typical time $t_{\rm edge}=\hbar/\Delta\epsilon_{\rm edge}$
tends to be infinity.
This means that for the adiabatic condition to be strictly satisfied at the edge
the pumping cycle $T$ must be infinite.
Fortunately, this condition will not be satisfied experimentally; 
the pumping cycle $T=T_{\rm exp}$ is well between the above
two time scales: $t_{\rm bulk} \ll T_{\rm exp}\ll t_{\rm edge}$,
i.e., the bulk is adiabatic, while the \cin{edges in  experiments are non adiabatic,
  that is, described by the } ``sudden'' \cin{approximation}.
Therefore, the jumps are not seen in the the experiments, even not in the numerical simulations of time evolution,
while the ``reconnection'' is justified; i.e., one can safely skip the jumps. 
This is because
in the regime of $T_{\rm exp}\ll t_{\rm edge}$,
the sudden approximation \crep{seems to be} {is} fully justified at the edge.
\footnote{\crep{Landau Lifshiz, ``Quantum Mechanics'', chap. xx}{A. Messiah, ``Quantum mechanics'', chap. XVII, $\S 7$.}}
The system behaves before and after the jump as if the jump \crep{did}{does} not exist.

Finally, let us recall that
\crepp{
  pumping is a many-body phenomenon in the sense
that it is essentially related to the change of occupancy of the edge states
in the course of the pumping cycle.}
      {the CM is only well-defined for an open system 
        although the pumped charge
        is described by the polarization of the bulk,
        which is compensated by the discontinuities due to the  edge states
        in the extreme adiabatic limit.
      }
Even though we may not see the jumps in the experiment,
{\it this part must underlie} for %a finite pumping 
the whole phenomenon to occur.

\ack
The authors are supported by KAKENHI:
15K05131 (KI), 15H0370001 (KI), 17H06138 (KI, TF, YH) and 16K13845 (YH).

\vspace{1cm}

\section*{References}
\bibliography{proc_LT28v2}

\end{document}